\documentclass[ ]{copernicus2}

\usepackage{color}
\usepackage[T1]{fontenc}

\begin{document}

\title{Olbertian partition function in scalar field theory
}

{\author[1,3]{R. A. Treumann}
\author[2]{W. Baumjohann$^a$}
\affil[1]{International Space Science Institute, Bern, Switzerland}
\affil[2]{Space Research Institute, Austrian Academy of Sciences, Graz, Austria}
\affil[3]{Geophysics Department, Ludwig-Maximilians-University Munich, Germany\protect\\
Correspondence to: Wolfgang.Baumjohann@oeaw.ac.at}

}

\runningtitle{Olbertian}

\runningauthor{R. A. Treumann \& W. Baumjohann}

\received{ }
\pubdiscuss{ } 
\revised{ }
\accepted{ }
\published{ }


\firstpage{1}

\maketitle

  

\noindent\textbf{Abstract}.-- 
{The Olbertian partition function is reformulated in terms of continuous (Abelian) fields described by the Landau-Ginzburg action, respectively Hamiltonian. In order do make some progress, the Gaussian approximation to the partition function is transformed into the Olbertian prior to adding the quartic Landau-Ginzburg term in the Hamiltonian. The final result is provided in the form of an expansion suitable for application of diagrammatic techniques once the nature of the field is given, i.e. once the field equations are written down such that the interactions can be formulated. } 

\section{Introduction}
Particle spectra in near-Earth space (for examples see \citep{christon1988,christon1989,christon1991}) as well as in cosmic rays very frequently exhibit power law tails at high energies  which since their introduction by Olbert \citep{olbert1967} have been interpreted as Olbert distributions ($\kappa$-distributions).\footnote{They were invented by Stan Olbert in 1966 and applied first in the unpublished PhD thesis of Binsack \citep{binsack1966} who acknowledged its suggestion by Olbert. (We thank C. Tsallis for kindly bringing this reference to our attention.) Olbert also suggested it to Vasyliunas whose paper \citep{vasyliunas1968} contained the first refereed, published and thus multiply cited version of the Olbert ($\kappa$) distribution.} Cosmic ray spectra in particular extend as power laws over many decades reminding of several ultra-relativistic Olbert distributions adding continuously up \citep{treumann2018}. Olbert distributions have  been inferred in plasma turbulence and many other occasions as for instance in front \citep{eastwood2005} and behind \citep{lucek2005} collisionless shocks \citep{balogh2013} as also, for example, in the Heliosphere and its Heliosheath \citep{fichtner2020} which may serve as the paradigm of a  stellar wind that is terminated by its interaction with the interstellar galactic medium. They were also derived in plasma wave-wave interaction theory \citep{hasegawa1985,yoon2005,yoon2006}. Physically they represent quasi-stationary states far from equilibrium \citep{scherer2020,livadiotis2013,treumann1999,treumann2004,domenech-garret2015}. To some degree they are related to Tsallis' thermostatistics \citep{tsallis1988}. We recently \citep{treumann2020} investigated their connection to Olbert's entropy.\footnote{Reviews of various definitions of entropy can be found in \citet{wehrl1978}, \citet{renyi1970}, \citet{tsallis1988}, \citet{kittel1980}.} Here we are interested in the role they might play in field theory which is the continuous version of the partition function \citep{zinn1989,kogut1974,parisi1988,reichl1980,aitchison1993,bailin1993,ryder1991}. We do not go into the definition what is meant by the term continuous. Fields are a basic concept of physics when many degrees of freedom come into play. In this case one refers to action and Hamiltonian densities which are distributed in space-time, and for the statistical interacting fields and distributed sources one refers to the field partition function from which the effects of the interaction between fields and particles can be deduced. In the following we derive the Olbertian partition function for scalar fields{,  the so-called generating functional}. One particularly interesting application {which could be anticipated} is in the cosmological theory of the early universe where phase transition are the rule \citep{kolb1989}. As it turns out, this is a highly non-trivial problem whose difficulties go substantially beyond those encountered with the Gibbsian partition function. Nevertheless we show, how the Olbertian field partition function can be constructed giving it an operational representation that is suitable for application. This is interesting in as far as the extra parameter $\kappa$ which is fundamental to Olbert theory provides an  external degree of freedom which may become useful in applications like renormalization and phase transition where convergence is obliterated.   

\section{Olbertian distribution -- a brief review}
{The Olbertian partition function (\ref{eq-gpart}) we are going to investigate in the next section is the Gibbs normalization factor of the Olbert distribution\footnote{In physics (and science in general) it is customary to assign the names of their inventors to theories or equations in order to be specific and make them indistinguishable. One speaks of the Boltzmann equation, Gibbs statistical mechanics, Tsallis statistics instead $q$-statistics, Heisenberg's relations, Einstein's theory, Feynman integrals and so on, unless the theory has its own indistinguishable name like QED, QCD, GRT and so on. The name $\kappa$-distribution is inappropriate. The letter $\kappa$ is nothing specific. The only appropriate name for the Olbert distribution and everything which follows from it would be Olbert's.}
\begin{equation}\label{eq-1}
p_\kappa(E_\alpha)=e^{-E_\alpha/E_c}\Big[1+\frac{\beta E_\alpha}{\kappa}\Big]^{-(\kappa+s)}
\end{equation}
with real $\kappa>0$ and $s$ some fixed number (for a recent account of the Olbert distribution and entropy see for instance \citep{treumann2020}). This distribution applies to finite temperature $\beta\ll\infty$ as we have discussed in previous work. Its limit for $\kappa\gg 1$ is the Boltzmann factor, as can easily be shown by applying l'Hospital's rule. Thus the Olbert distribution is a finite temperature distribution. At $\beta\to\infty$ large but still finite and $\kappa>s$ it is simple matter to show that the Olbert distribution becomes the low-temperature Boltzmann distribution $\sim\exp\big[-(1+s/\kappa)\beta E_\alpha\big]$ with unimportant modification factor which reduces the effective Boltzmann constant $k_B/(1+s/\kappa)$ a tiny bit.}

{For finite $\kappa$ this function is essentially a skewed Maxwellian (or Boltzmann function) with power law wings in momentum space (in energy space tails). It was widely discussed in the literature and applied to observed high energy tails on particle distribution  functions as well as formally, interpreting it as a statistical not a physical distribution, in statistical data analysis to demonstrate the existence of correlations which give rise to higher order moments like kurtosis. Sometimes it is claimed that the above probability (neglecting the exponential cut-off at energy $E_c\gg E_\alpha$) is the solution of a first order Pearson differential equation 
\begin{equation}
f'_\kappa(x)=-\frac{\kappa+s}{\kappa}\frac{f_\kappa(x)}{1+x}, \qquad x=\frac{\beta E_\alpha}{\kappa}
\end{equation}
which is obtained (in fact trivially) when expressing the first derivative of $p_\kappa(x)$ with respect to its argument through $p_\kappa(x)$.  This has no fundamental physical meaning however. Olbert distributions are stationary quasi-equilibria in the sense of statistical mechanics. All physical content is contained in the partition function given below. The index $\kappa$ accounts for the presence of internal correlation in the inner parameter space. It may in fact on its own be a function of momentum, energy, temperature, and some other fields like waves. In most application it is considered constant or at least a very slowly variable quantity which contains all the hidden information which is transferred in the unresolved internal space. Below we also consider it constant for all our purposes. It is for these reasons that the Olbert distribution in its conventional form applies to systems of finite temperature only, if $\kappa $ is not infinite. This is because classically one expects correlations to occur only at finite temperature. Quantum mechanical versions of the Olbert distribution require the assumption of non-ideal gases embedded into external force fields. Otherwise definition of a quantum mechanical analogue to the Olbert distribution so far is an only partially resolved task for bosons and anyons. It requires correlations and thus at low temperatures does probably not exist, because the only correlations available are those of vacuum fluctuations, which however are still controversial. Whether an Olbert distribution can exist at all for fermions is not known either even at finite temperature because it might be forbidden by the Pauli principle. Anyway, the problem of Olbert quantum distributions has not yet been settled satisfactorily possibly requiring a more sophisticated investigation of its low temperature properties and modifications.} 

{Distribution functions similar to the Olbert kind have been obtained in theory in various different ways. The original theoretical attempt to find a plausible explanation for them almost immediately referred to a Fokker-Planck approach. The Fokker-Planck equation  was naturally assigned to describe the formation of deformed non-stationary distributions by stochastic diffusion in phase space caused by electromagnetic fluctuations. It  requires prescription of a phase space diffusion coefficient and, to some extent, corresponds to quasilinear theory where under collisionless conditions first order wave-particle interactions scatter some of the particles into higher energy states while confining the lower energy bulk. This process, for an energy dependent diffusion coefficient, maintains the low energy particles in the Maxwellian,  while the unconfined higher energy particles run away into a skewed energetic tail that evolves with time. Fokker-Planck approaches are time dependent. One particular such diffusion coefficient is Coulomb scattering (Spitzer conductivity) which is energy dependent and thus may considered basic to scattering and power law tails. Coulomb scattering times are, however, very long such that it takes much time until an appreciable tail evolves. It is widely used for this property in cosmic ray physics where infinite time is available.  In order to obtain a stationary state, Fokker-Planck approaches require  continuous injection and some additional mechanism which cuts the distribution at some energy by removing energetic particles causing particle losses. Stationarity is achieved for instance when particles escape from the system or by imposing other particle sinks like for instance charge exchange with an atmosphere. }

{The Olbert distribution, on the other hand is a basic stationary quasi-thermodynamic state far from thermal equilibrium. It has been given a derivation from basic statistical mechanics where it can evolve under particular conditions, caused by the intrinsic structure of the distribution. It has been discussed in two forms from two different points of view, as thermostatistics \citep{tsallis1988} where an approximate form has been found, and in Lorentzian Gibbs-Boltzmann statistical mechanics \citep{treumann1999,treumann2014}. The latter leading directly to the Olbert distribution, the former constructing a similar distribution in suitable approximation. For a collection of other properties of the Olbert distribution the reader is directed to the extended literature on $\kappa$ distributions \citep[cf., e.g.,][and referenes therein]{livadiotis2013,scherer2020}.} 

{It is clear that the micro-physics of the formation of the Olbert distribution is contained in the free parameter $\kappa(\epsilon, \langle w\rangle, \beta)$ which itself can depend on particle energy $\epsilon$, temperature $\beta^{-1}$ or the underlying wave-particle interaction, i.e. the average energy $\langle w\rangle$ of the wave spectrum $w_k$. In fact,  in some cases theoretical approaches have derived functional expressions of $\kappa$ for electrons immersed into a photon bath \citep{hasegawa1985} and also in weak plasma turbulence \citep{yoon2005,yoon2006,yoon2018}, when electrons are subject to spontaneous, induced emission and absorption of Langmuir/ion-sound waves. In both cases the long-term limit provides the expected completely collisionless formation of the energetic tail and the deformation of the original Maxwellian distribution. One thus can, in a picture in that the system is in instantaneous equilibrium, the usual thermodynamic or statistical mechanical assumption, treat an Olbertian system as a thermodynamic state far from thermal equilibrium. Since the methods of thermodynamics and statistical mechanics are well-developed, this is particularly advantageous as it provides an analytical means and deep physical understanding of processes like phase transitions, heating, entropy formation and transport without the necessity to apply the restricted methods of numerical simulation as these produce figures in a limited parameter range which require subsequent physical interpretation. Or otherwise apply some approximations from nonlinear kinetic theory which quickly run into non-manageable complications when exceeding  quasilinear theory. In all approximately stationary cases subject just to long term variations the quasi-thermodynamic treatment far from thermal equilibrium provides useful information about the physics involved. This particularly true when the power law tail extends uniformly over more than just one decade in energy, frequency or wavenumber.}

\section{Gauss-Olbertian Theory}
The Olbertian partition function based on the Olbert distribution when normalizing it according to the Gibbs prescription is given by
\begin{equation}\label{eq-gpart}
Z_{O\kappa,s}(\beta)=\sum_\alpha e^{-E_\alpha/E_c}\Big[1+\frac{\beta E_\alpha}{\kappa}\Big]^{-(\kappa+s)}
\end{equation}
where $E_\alpha$ is energy in state $\alpha$, $\beta E_c\gg 1$ some high-energy cut-off \citep{lazar2020,scherer2017,scherer2020}, and $\kappa, s$ are exponents, of which $r$ is free to choose for satisfying thermodynamic needs. This partition function holds for discrete particles occupying the energy states $E_\alpha$. When the states are continuously distributed in space one switches to a field description. The restriction to space is no limitation, it simply describes that the field is continuously distributed with respect to some vector field $\mathbf{x}$.

\subsection{Olbertian partition function}
We here {for simplicity} generalize the Olbert partition function to given \emph{scalar} fields $\mathbf{\Phi}(\mathbf{x})$, which depend continuously on space $\mathbf{x}$, by introducing the functional integral
\begin{equation}\label{eq-z}
Z_{O\kappa,s}[J\,]\equiv\int\mathcal{D}\Phi\Big[1+\beta{H}[J\,]/\kappa\Big]^{-(\kappa+s)}
\end{equation}
The integration is over all realizations of the field $\Phi$ in physical space.\footnote{An idea first introduced by Einstein \citep{einstein1905} in his approach to the stochastic motion of molecules.} In field theory the partition function is conventionally called the ``generating functional''. The Hamiltonian $H=H_{LG}$ we will investigate is the Landau-Ginzburg Hamiltonian
\begin{equation}\label{eq-4}
H_{LG}=\int d\mathbf{x}\,\mathcal{H}_{LG}\big[\Phi(\mathbf{x}),\mathbf{J}(\mathbf{x})\big]
\end{equation}
with Landau-Ginzburg Hamiltonian density
\begin{equation}\label{eq-5}
\mathcal{H}_{LG}\big[\mathbf{\Phi}(\mathbf{x}), \mathbf{J}\,\big]\equiv \frac{\alpha^2}{2}|\nabla\Phi|^2+\frac{\mu^2}{2}|\Phi|^2+\frac{\lambda}{4!}|\Phi|^4-\mathbf{J}\cdot\mathbf{\Phi}
\end{equation}
{originally introduced in superconductivity theory where $|\Phi|$ is some order parameter, in fact the normalized density of electron pairs. It was soon realized \citep{nambu1960,nambu1961,anderson1963} to be of substantial importance in field theory of elementary particles as well as in spontaneous symmetry breaking.} We include an external source term $\mathbf{J\cdot}\mathbf{\Phi}$ which results from the interaction of $\mathbf{J}$ with the field $\mathbf{\Phi}$. Since the field is scalar this interaction is restricted to the source $\mathbf{J}$ and field $\mathbf{\Phi}$ being parallel, for example in a magnetic field $\mathbf{B}$ one would have $\mathbf{J}=\mathbf{B}$, {which is the case in superconductivity theory where the stationary magnetic field is external and experiences the Meissner effect.} The interaction term in the dimensionless expression $\beta H_{LG}$ then becomes $\beta\mathbf{B}\cdot\mathbf{\Phi}$.  As usual the powers of the dimensionality $L^\nu$ of the various quantities are $[\mathcal{H}_{LG},\Phi,\alpha,\mu^2,\lambda]=(-d,-d/2,1,0,d)$. The above partition function and Hamiltonian are rather complicated. Note that we did not include any cut-off {like in particle  theory} as this will come in below in a quite natural way for the fields as a cut-off at small, ultimately molecular scales if not imposed by physics at larger scales already. If the dynamic equations governing the field are required, these are obtained from the action principle based on the above Hamiltonian by the well-known method of variation which by Noether's theorem also yields the conserved quantities, so-called Noether currents. 

{In the above expressions $\mathcal{H}_{LG}$ is the field Hamiltonian density which seems to be a function of $\mathbf{x}$ only apparently lacking any momentum dependence. This is, however, not the case. The transformation from particle to classical spatially dependent continuous scalar field variables, here the only interesting case, is done via the Lagrange formulation which yields the equivalence $\mathbf{x}\to\Phi(\mathbf{x})$ for the spatial variable, and $\mathbf{p}\to \nabla\Phi(\mathbf{x})$ for the momentum $\mathbf{p}$. (In vector fields, which we are not dealing with here, this equivalence becomes slightly more complicated.) Hence results the functional dependence of $\mathcal{H}_{LG}\big[\Phi(\mathbf{x})\big]$, with field function containing the spatial dependence which appears in both, the field and its momentum, its spatial derivative. It is the gradient term in the Landau-Ginzburg Hamiltonian that, with appropriate dimensional coefficient, takes care of the field momentum which in the particle picture would read $\mathbf{p}^2/2m$ for nonrelativistic particles. In the theory developed below it generates an important wavenumber dependence. }

{We also note that the Hamiltonian is an energy density. Since energy is additive, one can simply add any interacting external field as, for instance, the energy density of the electromagnetic field $F_{\mu\nu}F^{\mu\nu}$, with $F_{\mu\nu}$ the electromagnetic field tensor. Including it in the theory below is simple matter and for our purposes here not required as it  would unreasonably blow up the expressions. It would also require replacing the gradient with the covariant derivative $\nabla\to\nabla-ie\mathbf{A}(\mathbf{x})$ with $\mathbf{A}(\mathbf{x})$ the vector potential, which in this case couples the scalar and vector fields.}

As usual the Landau-Ginzburg Hamiltonian, even though it is an expansion just up to second order in the modulus of the field $|\Phi|^2$ containing most of the relevant physics, in particular spontaneous symmetry breaking and phase transition, provides considerable difficulty when introduced in the partition function. These difficulties are twofold. In general the integration over the various realizations of the field set a barrier for the treatment. This corresponds, however, to the usual difficulty appearing already in the particle picture. The second and more serious difficulty arises with the symmetry breaking quartic (4th order) term in the field. One therefore needs to investigate the partition function in two steps with the first step remaining in the realms of the ordinary Gaussian form of the Landau-Ginzburg Hamiltonian as used in Landau superfluidity, dropping the quartic term, which will be included in a second step when already having given the Gaussian partition function its suitable representation.

\subsection{Gauss-Hamiltonian approach}
So, in order to proceed we provisionally drop the quartic term  in the absence of any external field and self-interaction, and consider the Gauss-Olbertian partition function $Z_{GO}[J]$ with Hamiltonian density
\begin{equation}\label{eq-6}
\mathcal{H}_{G}\equiv \frac{1}{2}\alpha^2|\nabla\Phi|^2+\frac{1}{2}\mu^2|\Phi|^2-J\Phi
\end{equation}
This Hamiltonian is assumed to hold for any not particularly specified field $\Phi(\mathbf{x})$. Hence, in order not having to worry about the dimension of the field we assume that it is normalized to either its mean, minimum or expectation value. {In this case $\Phi(\mathbf{x})$ is dimensionless, while $\mathcal{H}_{G}$ maintains its physical dimension as energy density which we can take out as a common factor. Thus $\mu, \lambda, J$ have no dimension from here on, while the dimension of $\alpha$ is length (or $k^{-1}$) such that $\alpha k$ is dimensionless.} Following standard methods, the fields and source terms can be Fourier transformed, defining wavenumber space $d\mathbf{k}/(2\pi)^d\to d\mathbf{k}$ and
\begin{eqnarray}\label{eq-7}
\Big(\Phi(\mathbf{x}), J(\mathbf{x})\Big)&=&\int^{k_c}d\mathbf{k}\,e^{i\mathbf{k\cdot x}}\Big(\phi(\mathbf{k}),j(\mathbf{k})\Big)\\
\Big(\phi(-\mathbf{k}),j(-\mathbf{k})\Big)&=&\Big(\phi^*(\mathbf{k}),j^*(\mathbf{k})\Big)
\end{eqnarray}
where the wavenumber $k_c$ is an upper cut-off which takes care of cutting the spectrum of fluctuations at the molecular scale. This overwrites the necessity of including the high-energy cut-off in the above Olbertian (particle) partition function. The Gaussian reduction $H_G$ of the Landau-Ginzburg Hamitonian is the $\mathbf{x}$-integral of $\mathcal{H}_G$ when using the Fourier transformed quantities which yields \citep{amit1984,binney1999,parisi1988,reichl1980,zinn1989} its well-known representation
\begin{eqnarray}\label{eq-8}
H_G&=&L^{-d}\sum_{ k<k_c}(\alpha^2k^2+\mu^2)[\phi^2_{r\mathbf{k}}+\phi^2_{i\mathbf{k}}] \nonumber\\
&+&2(\phi_{r\mathbf{k}}j_{r\mathbf{k}}+\phi_{i\mathbf{k}} j_{i\mathbf{k}})\ \ \equiv \sum_{ k<k_c} H_{Gk}
\end{eqnarray}
Indices $r,i$ design  real and imaginary parts of the Fourier amplitudes of the field and source. The real and imaginary parts thus decouple. In this form the Hamiltonian holds for any interaction source field $J$, and the Gauss-Olbertian partition function Eq.\,(\ref{eq-z}) becomes
\begin{equation}\label{eq-hg0}
Z_{GO\kappa,s}[J\,]\equiv\int d\phi_{rk}d\phi_{ik}\Big[1+\beta {H}_G[J]/\kappa\Big]^{-(\kappa+s)}
\end{equation}
where the integration is with respect to the decoupled real and imaginary fields. The unity in the bracket can in principle be absorbed into the Hamiltonian which is a sum. It is however rather inconvenient to work with this form because handling the finite sum in the denominator at this stage prevents any further progress unless it can be summed up, which however requires a functional form of the functions $\phi_k$ and $j_k$. Therefore a different strategy is needed. 

For any external source $\mathbf{J}$ the Hamiltonian {(\ref{eq-8})} can also be transformed into a more symmetric form by re-shuffling the real and imaginary parts and completion to squares. This yields
\begin{equation}\label{eq-hg}
H_G=L^{-d}\sum_{ k<k_c}(\alpha^2k^2+\mu^2)|\phi'_{{k}}|^2-\frac{|j_{k}|^2}{\alpha^2k^2+\mu^2}
\end{equation}
where the new field is $$\phi'_{r,ik}=\phi_{r,ik}+j_{r,ik}/(\alpha^2k^2+\mu^2)$$ {which maintains the symmetry of the Hamiltonian}. The field integration in the partition function is then with respect to $\frac{1}{2}\int \,d|\phi'_k|^2d\theta_k\, e^{i\theta_k}$, where $\tan\theta_k=\phi'_{ik}/\phi'_{rk}$. In the particular case of an isotropic field like, for instance, the scalar Higgs field $\Phi_H$, the phase integration becomes trivial, reducing to a factor $2\pi i$. The above integral then assumes a slightly simpler form. 

\subsection{Gauss-Olbertian Partition Function}
Let us return to the Gibbs-Boltzmann partition function $$Z_{GB}=\sum_\alpha\exp(-\beta E_\alpha)$$ writing it in terms of the field $\Phi$ and use its Gaussian result which is widely applied in {finite temperature field theory \citep{amit1984,binney1999}}. In that case the simplicity of the exponential form of the Gibbs-Gaussian partition function after Fourier transformation allows to factorize the resulting integral  
\begin{equation}\label{eq-9}
Z_{GB}[J\,]=\exp\bigg[\frac{1}{2}\int^{k_c}d\mathbf{k}\frac{j_kj_{-k}}{\alpha^2k^2+\mu^2}\bigg]Z_{GB}[0]
\end{equation}
with the definition
\begin{equation}\label{eq-10}
Z_{GB}[0]=\exp\bigg[-\frac{V}{2}\int^{k_c}d\mathbf{k}\log(\alpha^2k^2+\mu^2)\bigg]
\end{equation}
$V$ is the normalizing volume (for instance of a large box). These expressions transform directly into an explicit form of the Gauss-Olbertian partition function for the fields
\begin{eqnarray}\label{eq-11}
Z_{GO}[0]&=&\bigg[1+\frac{V}{2\kappa}\int^{k_c}d\mathbf{k}\log(\alpha^2k^2+\mu^2)\bigg]^{-\kappa-s}\nonumber\\
Z_{GO}[J\,]&=&\bigg\{1-\frac{1}{2\kappa}\int^{k_c}d\mathbf{k}\frac{j_kj_{-k}}{\alpha^2k^2+\mu^2}\bigg\}^{-\kappa-s}Z_{GO}[0]
\end{eqnarray}
From here one observes that in the absence of an external source field the function in the second line reduces to one, and only the zero order partition function remains. 
The free energy then follows immediately from 
\begin{eqnarray}\label{eq-12}
F_{GO}[J\,]&=&-\beta^{-1}\log\,Z_{GO}[J\,]\\
&=&\frac{\kappa+s}{\beta}\log\bigg\{\bigg[1+\frac{V}{2\kappa}\int^{k_c}d\mathbf{k}\log(\alpha^2k^2+\mu^2)\bigg]\nonumber\\
&&\qquad\qquad\!\times\,\bigg[1-\frac{1}{2\kappa}\int^{k_c}d\mathbf{k}\frac{j_kj_{-k}}{\alpha^2k^2+\mu^2}\bigg]\bigg\}
\end{eqnarray}
a form that can be used to derive further quantities of thermodynamic interest like the energy and the specific heat. The form given is not yet suited for calculating correlation functions, because it contains the Fourier transformed sources. Returning in the partition function to the original sources $J$ via the inverse transformation and using the particular form of the Dirac function
\begin{equation}\label{eq-13}
\delta(\mathbf{x}_1-\mathbf{x}_2)\equiv \int^{k_c^{-1}}\frac{d\mathbf{k}}{\alpha^2k^2+\mu^2}e^{i\mathbf{k}\cdot(\mathbf{x}_1-\mathbf{x}_2)}, \quad\mu^2>0
\end{equation}
{used in field theory at finite temperatures which results from the two-point correlation function, whose properties we give below}, the source integral can be rewritten in a form suitable for taking derivatives with respect to $J$
\begin{equation}\label{eq-14}
\int^{k_c}\frac{d\mathbf{k}\,j_kj_{-k}}{\alpha^2k^2+\mu^2}=\int d\mathbf{x}_1d\mathbf{x}_2J(\mathbf{x}_1)\delta(\mathbf{x}_1-\mathbf{x}_2)J(\mathbf{x}_2)
\end{equation}
This is to be used in the partition function $Z_{GO}[J]$ {Eq.\,(\ref{eq-11})} when calculating any $n$th order correlations 
\begin{equation}\label{eq-15}
C^{(n)}(x_1\dots x_n,J)=\prod_i \frac{\delta}{\delta J(x_i)}\log Z_{GO}[J(x_1\dots x_n)]
\end{equation}
{This last expression is the definition of the correlation functions. Forming the one-point correlation function yields $C^{(1)}\equiv\langle\Phi\rangle=\int d\mathbf{x}_2\delta(\mathbf{x}_1-\mathbf{x}_2)J(\mathbf{x}_2)$ just the average of the field. The two-point correlation function is equal to the above Dirac function:
\begin{eqnarray}\label{eq-16}
C^{(2)}(\mathbf{x}_1,\mathbf{x}_2)&=&\langle[\Phi(\mathbf{x}_1)-\langle\Phi(\mathbf{x}_1)\rangle][\Phi(\mathbf{x}_2)-\langle\Phi(\mathbf{x}_2)\rangle]\rangle\nonumber\\ 
&=& \langle\Delta\Phi(\mathbf{x}_1)\Delta\Phi(\mathbf{x}_2)\rangle \ \ = \ \ \delta(\mathbf{x}_1-\mathbf{x}_2)
\end{eqnarray}
showing that the Dirac function (\ref{eq-13}) is nothing else but the representation of the two-point correlation function (\ref{eq-16}) which is an identity and of practical use in the following. It is otherwise easy to show that (\ref{eq-13}) is indeed a representation of the Dirac function yielding the self-correlation of the field. Higher order correlations do not exist in this approximation but may indeed occur in the more precise Olbert theory below.} 

{We briefly note as a side remark that explicit calculation confirms that Eq.\,(\ref{eq-13}) is indeed a Dirac function. Since its denominator is symmetric in $k$, the integral can be analytically continued into the negative domain with the poles shifted along one of the axes by the amount $\pm\mu/\alpha$, depending on the sign of $\mu^2$ the real or imaginary axis. In our  $\mu^2>0$ case the pole are symmetrical along the imaginary axis. Thus they are at $k\ll k_c$ and the cut off plays no role. For these purposes it can be assumed to be at infinity. Splitting the three dimensional integral produces the sum of an ordinary Dirac function plus two singular integrals with simple poles whose contributions can easily be obtained when appropriately choosing the paths of phase integration around the poles, in which case their contributions vanish. This maintains the character of the Dirac function as a distribution in the dominant first integral and is applied in Eq.\,(\ref{eq-14}) when re-transforming into configuration space as, is needed when intending application of the well-known field-theoretical diagrammatic order-by-order interaction method through functional derivatives \citep{amit1984,binney1999}.}      

The above approximate Gauss-Olbertians are nevertheless in a still fairly inconvenient form. This can be made more explicit by writing them as exponentials. For instance we have
\begin{equation}\label{eq-17}
Z_{GO}[0]=\exp\bigg[{-(\kappa+s)}\log\Big(1+\frac{V}{2\kappa}\int^{k_c}d\mathbf{k}\log(\alpha^2k^2+\mu^2\Big)\bigg]
\end{equation}
Expanding  both the logarithm in the argument of the exponential and subsequently the exponential itself produces {in Eq.\,(\ref{eq-11})}
\begin{equation}\label{eq-18}
Z_{GO}[0]=\prod_{n=1}^\infty\sum_{m=0}^n\frac{1}{m!}\Big(\frac{\kappa+s}{n}\Big)^m\Big(-\frac{V}{2\kappa}\int^{k_c}d\mathbf{k}\log(\alpha^2k^2+\mu^2\Big)^{nm}
\end{equation}
and correspondingly
\begin{eqnarray}\label{eq-19}
Z_{GO}[J\,]&=&\prod_{n=1}^\infty\sum_{m=0}^n\bigg[-\frac{1}{m!}\Big(\frac{\kappa+s}{n}\Big)^m\nonumber\\
&\times&\bigg\{\frac{1}{2\kappa}\int^{k_c}d\mathbf{k}\frac{j_kj_{-k}}{\alpha^2k^2+\mu^2}\bigg\}^{nm}\bigg]\ Z_{GO}[0]
\end{eqnarray}
This, with the above inverse Fourier representation, assumes the final form
\begin{eqnarray}\label{eq-20}
Z_{GO}[J\,]&=&\prod_{n=1}^\infty\sum_{m=0}^n\bigg[-\frac{1}{m!}\Big(\frac{\kappa+s}{n}\Big)^m\\
&\times&\bigg\{\frac{1}{2\kappa}\int d\mathbf{x}_1d\mathbf{x}_2J(\mathbf{x}_1)\delta(\mathbf{x}_1-\mathbf{x}_2)J(\mathbf{x}_2)\bigg\}^{nm}\bigg]\ Z_{GO}[0]\nonumber
\end{eqnarray}
as an approximate representation of the Gauss-Olbertian partitian function. Though this is just a simplified form which we below will make more precise, it will in most cases be sufficient in applications. 

\subsection{Rigorous derivation}
The final last forms of the Gauss-Olbertian partition function provide a feasible way to go but are not completely satisfactory as the transformation from the Gibbs-Gaussian to the Gauss-Olbertian is done on a late stage. It can be taken as a lowest order approximation to a more precise formulation of the Olbertian field-partition function whose derivation is attempted in the following.   
We therefore tentatively return to the formal definition of the Gauss-Olbertian partition function (\ref{eq-hg0}) through the Hamiltonian $\mathcal{H}_G$ and attempt a different approach. In fact, the partition function can also be written
\begin{equation}\label{eq-21}
Z_{GO\kappa,s}[J\,]\equiv\int d\phi_{rk}d\phi_{ik}e^{-(\kappa+s)\log(1+\beta {H}_G[J]/\kappa)}
\end{equation}  
The logarithm in the exponential can then be expanded if taking care that the Hamiltonian has been appropriately normalized. This yields
\begin{eqnarray}\label{eq-22}
Z_{GO\kappa,s}[J]&\equiv&\int d\phi_{rk}d\phi_{ik}\exp\Big[\sum_{n=1}\frac{\kappa+s}{n}\Big(-\frac{\beta H_G}{\kappa}\Big)^n\Big]\nonumber\\
&=&\prod_{n=1}\int d\phi_{rk}d\phi_{ik}\exp\Big[\frac{\kappa+s}{n}\Big(-\frac{\beta H_G}{\kappa}\Big)^n\Big]
\end{eqnarray} 
However, $H_G$ is itself a sum. So in the argument of the exponential one has the power of a sum. In order to avoid a much bigger complication we find that the second of these versions is the most convenient one when we expand the exponential in powers. This yields the $m$-ordered expression
\begin{equation}\label{eq-23}
Z_{GO\kappa,s}[J]= \prod_n\sum_m^n\Big(-\frac{\beta}{\kappa}\Big)^{nm}\Big(\frac{\kappa+s}{n}\Big)^m\int d\phi_{rk}d\phi_{ik}\Big(\sum_{k<k_c}H_{Gk}\Big)^{nm}
\end{equation}
where $H_{Gk}$ has been defined already through Fourier transformed quantities {as the summand in Eq.\,(\ref{eq-hg})}. Though this is more precise, it is rather involved and therefore inconvenient for further exploration.  

We therefore return to the Hamiltonian $H_G$ which is a sum of two terms. It can in principle be expanded into a binomial series
\begin{equation}\label{eq-24}
H_G^n=\sum_{\ell=0}^n\Big({n\atop \ell}\Big)H_{G\phi}^\ell H_{GJ}^{n-\ell}
\end{equation}
where the two forms of $H_G$ are the two terms in the definition equation (\ref{eq-hg}). It appears in the argument of the exponential function, which allows to write the partition function as a double product
\begin{equation}\label{eq-hg3}
Z_{GO\kappa,s}=\prod_{n=1}^\infty\prod_{\ell=0}^n\int d\phi_{rk}d\phi_{ik}\exp\Big[\frac{\kappa+s}{n}\Big(-\frac{\beta}{\kappa}\Big)^n\Big({n\atop\ell}\Big)H_{G\phi}^\ell H_{GJ}^{n-\ell}\Big]
\end{equation}
The two Hamiltonian functions are themselves sums with respect to $k$, here being multiplied with each other term by term such that their product becomes
\begin{equation}\label{eq-25}
H_{G\phi}^\ell H_{GJ}^{n-\ell}=\Big[\sum_{k_1}^{k_c}H_{G\phi,k_1}\Big]^\ell \Big[\sum_{k_2}^{k_c}H_{GJ,k_2}\Big]^{n-\ell}
\end{equation}
each sum consisting of two further sum terms. Only the second of these two factors depends on the source function $J$. Hence, if we expand it again into a binomial series leaving the first sum in its compact form, we find
\begin{eqnarray}\label{eq-26}
\Big[\sum_{k_2}^{k_c}H_{GJ,k_2}\Big]^{n-\ell}&=&\big(2L^{-d}\big)^{n-\ell}\sum_{p=0}^{n-\ell}\Big({n-\ell\atop p}\Big)\nonumber\\
&\times&\Big(\sum_{k_2}\phi_{rk}j_{rk}\Big)^p\Big(\sum_{k_2}\phi_{ik}j_{ik}\Big)^{n-\ell-p}
\end{eqnarray}
Again transforming the binomial sum into a product we arrive at
\begin{eqnarray}\label{eq-27}
Z_{GO\kappa,s}[J\,]&=&\prod_{n=1}^\infty\prod_{\ell=0}^n\prod_p^{n-\ell}\int d\phi_{rk}d\phi_{ik}\nonumber\\
&\times&\exp\Big[\frac{\kappa+s}{n}\Big(-\frac{\beta}{\kappa}\Big)^n\big(2L^{-d}\big)^{n-\ell}\Big({n\atop\ell}\Big)H_{G\phi}^\ell\\
&\times &\Big(\sum_{k_2}\phi_{rk}j_{rk}\Big)^p\Big(\sum_{k_2}\phi_{ik}j_{ik}\Big)^{n-\ell-p}\Big]\nonumber
\end{eqnarray}
The explicit dependence on $J$ is now contained only in the two factors in the last line. This expression still contains the sums over wave numbers which in a continuum representation can be replaced by integrals by reintroducing the inverse Fourier transform in these terms and using the Dirac function as given above. Without further assumptions about the field we are, however, stuck at this point. 

\subsection{Isotropic field}
The latter difficulty can be circumvented when assuming that we are dealing with an isotropic field. In this case the Hamiltonian, given in (\ref{eq-hg}), separates the source dependence out, and the source dependent Hamiltonian becomes
\begin{eqnarray}\label{eq-28}
H_{GJ}&=&-L^{-d}\sum_k^{k_c}\frac{j_kj_{-k}}{\alpha^2k^2+\mu^2}\nonumber\\
&\to&\int d\mathbf{x}_1 d\mathbf{x}_2J(\mathbf{x}_1)\delta(\mathbf{x}_1-\mathbf{x}_2)J(\mathbf{x}_2)
\end{eqnarray}
an expression that can directly be introduced in (\ref{eq-hg3}) since now the summand in this Hamiltonian is a single term such that the $p$-product disappears and the partition function assumes the simpler version
\begin{eqnarray}\label{eq-29}
Z_{GO\kappa,s}&\propto&\prod_{n=1}^\infty\prod_{\ell=0}^n\int \phi'_{k}d\phi'_{k}\exp\Big[\frac{\kappa+s}{n}\Big(-\frac{\beta}{\kappa}\Big)^n\Big({n\atop\ell}\Big)H_{G\phi'}^\ell\nonumber\\
&\times& \Big\{\int d\mathbf{x}_1 d\mathbf{x}_2J(\mathbf{x}_1)\delta(\mathbf{x}_1-\mathbf{x}_2)J(\mathbf{x}_2)\Big\}^{n-\ell}\Big]
\end{eqnarray}
where unimportant constant factors in front of the partition function have been suppressed. The field dependent Hamiltonian $H_{G\phi}$ is left untouched in this form. It is to be integrated over all realizations of the transformed field $\phi'_k$. This integration does not affect the integral in the last line. To separate this out one, as a final step at this stage, expands the exponential into a power series. This yields ultimately
\begin{eqnarray}\label{eq-zgfin}
Z_{GO\kappa,s}&\propto&\prod_{n=1}^\infty\prod_{\ell=0}^n\sum_{q=0}^\ell\frac{1}{q!}\Big[\frac{\kappa+s}{n}\Big({n\atop\ell}\Big)\Big]^q\Big(-\frac{\beta}{\kappa}\Big)^{nq}\nonumber\\
&\times& \Big\{\int d\mathbf{x}_1 d\mathbf{x}_2J(\mathbf{x}_1)\delta(\mathbf{x}_1-\mathbf{x}_2)J(\mathbf{x}_2)\Big\}^{q(n-\ell)}\\
&\times&\Big\{\int \phi'_{k}d\phi'_{k}H_{G\phi'}\Big\}^{q\ell}\nonumber
\end{eqnarray}
as the wanted form of the Gauss-Olbertian partition function. Once the integration with respect to $\phi'$ has been performed, it provides the correlation functions on each order by functional derivation with respect to $J$ in the Gaussian approximation. This integration with respect to the field can be done in the last term yielding
\begin{eqnarray}\label{eq-30}
\int \phi'_{k}d\phi'_{k}H_{G\phi'}&=& \frac{V}{4}\int^{k_c} d\mathbf{k} (\alpha^2k^2+\mu^2)|\phi'_k|^4\\ 
|\phi'_k|^2&=&\Big|\phi_k+\frac{j_k}{(\alpha^2k^2+\mu^2)}\Big|^2
\end{eqnarray}
where the sum over wave numbers has been replaced by the $k$-integration. One may note that $\phi'_k$ in this expression depends on the source which unavoidably introduces mixed field-source terms.

In this form we have obtained the final form of the Gauss-Olbertian partition function. It does not yet account for the quartic term in the Landau-Ginzburg Hamiltonian and therefore neither contains self-interactions of the field, nor spontaneous symmetry breaking which was the big progress and success in Landau-Ginzburg theory, nor does it account for the effects of the external source on symmetry breaking. It does, however contain Gaussian phase transitions. Nevertheless in its Gaussian form it is already subject to application of the diagrammatic technique using Feynman diagrams term by term. Still the last mixed integral term provides complications even here. As one observes, the form given to it here suggests that there is a huge number of terms which contribute, even though they contribute ordered by increasing power.  

In order to complete the theory to include spontaneous symmetry breaking on the level of the Olbertian partition function, one needs to refer to the quartic term in the Landau-Ginzburg Hamiltonian. This is the content of the next section.

\section{Landau-Ginzburg-Olbertian Theory}
So far we derived the Gauss-Olbertian partition function which, from the point of view of Landau-Ginzburg theory can be considered as the Gaussian approximation to the full Hamiltonian. In non-Olbertian field theory the Gibbs partition function is the usual exponential, and the quartic term in the Hamiltonian enters through the exponential in the integrand. It can be easily factorized such that the partition function is simply multiplied by an additional function however complicated that function appears. Expanding this exponential then gives an infinite series of correlations which are subject to Feynman representations of the hierarchy of interactions.  In Olbertian theory this is not anymore possible in this simple way, unless one chooses to simplify the theory substantially. The full Hamiltonian entering into (\ref{eq-z}) is
\begin{equation}\label{eq-31}
H_{LG}=H_{G}+\frac{\lambda}{4!}|\Phi|^4
\end{equation}
such that we can write
\begin{equation}\label{eq-zgen}
Z_{GO\kappa,s}[J]\equiv\int \mathcal{D}\phi\, e^{-(\kappa+s)\log(1+\beta {H}_G[J]/\kappa+\beta\lambda|\Phi|^4/4!\kappa)}
\end{equation}
Now, $\lambda<1$ should be a small expansion coefficient. So far we developed the as far as possible complete theory of the part containing just $H_G$. The simplest extension to $|\Phi|^4$ theory would be to assume that the term containing the quartic is an exponential. In that case the way would be to multiply the above final Gaussian result by the expansion of the quartic exponential and to write the Olbertian-Landau-Ginzburg partition function
\begin{equation}\label{eq-32}
Z_{OLG}[J\,]\propto \sum_p\frac{1}{p!}\Big\langle\Big[-\frac{\beta\lambda}{4!}\int d\mathbf{z}\Phi^4\Big]^p\Big\rangle_G\,Z_{GO\kappa,s}[J\,]
\end{equation}
(with dummy integration variable $\mathbf{z}=(\mathbf{x}_1,\mathbf{x}_2\dots)$). Here the subscript $G$ on the angular brackets indicates that we take the quartic term just in the Gaussian approximation, interpreting it as an operator which acts on the Gauss-Olbertian partition function explicated in (\ref{eq-zgfin}). The explicit form of this expression is found  in the literature [see e.g.\citep{amit1984,binney1999}] as an expansion in terms of functional derivatives
\begin{equation}\label{eq-33}
Z_{OLG}[J\,]\propto \sum_{p=0}^\infty\frac{1}{p!}\Big[-\frac{\beta\lambda}{4!}\int d\mathbf{z}\frac{\delta^4}{\delta J^4(\mathbf{z})}\Big]^pZ_{GO\kappa,s}[J\,]
\end{equation}
with the functional derivatives of order $4p$ acting on the Gauss-Olbertian partition function. This is a complicated though feasible result which can be taken as an approximation that cares already for some of the peculiarities of the Olbertian. Since the sum is but the expansion of an exponential, a slightly more precise approximation is to reinterpret it as an Olbertian writing
\begin{equation}\label{eq-34}
Z_{OLG}[J\,]\propto\Big[1+ \frac{\beta\lambda}{4!\kappa}\int d\mathbf{z}\frac{\delta^4}{\delta J^4(\mathbf{z})}\Big]^{-\kappa-s}Z_{GO\kappa,s}[J\,]
\end{equation}
where the exponential has been replaced by an Olbertian function. 

\subsection{Landau-Ginzburg-Olbertian}
The last expression can, in the usual way, be raised to an exponential which now retains the Olbertian functional form in the logarithm
\begin{eqnarray}\label{eq-35}
Z_{OLG}[J\,]&\propto&\exp\bigg\{-(\kappa+s)\log\Big[1+ \frac{\beta\lambda}{4!\kappa}\int d\mathbf{z}\frac{\delta^4}{\delta J^4(\mathbf{z})}\Big]\bigg\}\nonumber\\
&\times&\ Z_{GO\kappa,s}[J\,]
\end{eqnarray}
Expanding the logarithm in the argument of the exponential we arrive at another infinite product of exponentials
\begin{eqnarray}\label{eq-36}
Z_{OLG}[J\,]&\propto&\prod_{a=1}^\infty\exp\bigg\{(-1)^a\frac{\kappa+s}{a}\bigg[\frac{\beta\lambda}{4!\kappa}\int d\mathbf{z}\frac{\delta^4}{\delta J^4(\mathbf{z})}\bigg]^a\bigg\}\nonumber\\
&\times&\ Z_{GO,\kappa,s}[J\,]
\end{eqnarray}
The exponential can finally be expanded to give
\begin{eqnarray}\label{eq-olg}
Z_{OLG}[J\,]&\propto&\prod_{a=1}^\infty\sum_{b=0}^\infty\frac{(-1)^{ab}}{b!}\Big(\frac{\kappa+s}{a}\Big)^b\bigg[\frac{\beta\lambda}{4!\kappa}\int d\mathbf{z}\frac{\delta^4}{\delta J^4(\mathbf{z})}\bigg]^{ab}\nonumber\\
&\times& \ Z_{GO\kappa,s}[J\,]
\end{eqnarray}
This version of the Olbertian partition function contains the main features of  Olbertian theory and is thus a valid approximation when applied to scalar fields in Landau-Ginzburg theory.

\subsection{Validation}
The way of how to arrive at this result is now quite clear. In an even more precise theory than that given here one returns to the original Landau-Ginzburg Hamiltonian and repeats the sequence of steps. We briefly sketch how this is done. To this purpose one returns to (\ref{eq-zgen}), an expression which in the argument of the exponential contains the Gauss-Hamiltonian plus the Landau-Ginzburg term. The logarithm can advantageously be split into a sum of logarithms
\begin{eqnarray}\label{eq-37}
\log(1&+&\beta {H}_G[J]/\kappa+\beta\lambda|\Phi|^4/4!\kappa)=\log(1+\beta H_g/\kappa)\nonumber\\
&+&\ \log\Big[1+\beta\lambda|\Phi|^4/4!\kappa(1+\beta H_g/\kappa)\Big]
\end{eqnarray}
When inserted into the partition function and evaluated, the first term on the right leads to the above expression $Z_{GO\kappa,s}[J\,]$. It multiplies the exponential that contains the quartic term in the second logarithm. If we here  neglect the Hamiltonian $H_G$ in the denominator of the argument,  we just find the result given in (\ref{eq-olg}) which shows precisely which approximation has led to it. Writing it instead in the form
\begin{eqnarray}\label{eq-38}
\log\Big[1&+&\beta\lambda|\Phi|^4/4!\kappa(1+\beta H_g/\kappa)\Big]=\log\Big[1+\beta\lambda|\Phi|^4/4!\kappa\Big]\nonumber\\
&+&\log\Big[1+4!\kappa/(1+\beta H_g/\kappa)\beta\lambda|\Phi|^4\Big]
\end{eqnarray}
shows the effect of the Gaussian Hamiltonian just as a higher order correction factor. Thus one observes that the final result given above is the best available version of the Landau-Ginzburg-Olbertian partition function in a form which is suitable for application. Neglect of the second term on the right is sufficient justification for our approximation made in the former subsection to arrive at $Z_{OLG}[J\,]$ in (\ref{eq-olg}). Still it is free to choose either the approximate Gauss-Olbertian $Z_{GO}$ or its more precise though substantially more involved version $Z_{GO\kappa,s}$ as the basic Gauss-Olbertian in Landau-Ginzburg-Olbert field theory at finite temperature $ \beta\ll\infty$ and finite $\kappa$.

\section{Summary}
Field partition functions play a substantial role in field theory when interacting fields are under scrutiny. In particular, the partition function is the key to the identification of phase transitions on the one hand, on the other in managing renormalization \citep{amit1984,kogut1974} and elimination of divergences. Since Olbertian distribution functions have turned out to apparently building up frequently in nature where in the particle realm they can be treated by application of the partition function, it seems to make sense to investigate whether they can be reformulated as well in field theory. This has been done in the present note where it has been shown that, with some substantial modifications Olbertian field partition functions can be formulated and brought into a form which is suitable for application of diagrammatic techniques like Feynman diagrams if only the interactions can be identified. One can imagine that in various applications the Olbertian partition function may become useful. This may happen in systems which turn out to exhibit non-Gaussian behaviour. Classically such cases are familiar from turbulence theory where power laws are the rule thinking of, for instance, Kolmogorov spectra. Other candidates for application are the very early universe and phase transitions therein. Here we just provided the framework for application given in the formulation of the elaborated though convenient mathematical  structure of the Olbertian partition function.   

\section*{Acknowledgments}
 This work was part of a brief Visiting Scientist Programme at the International Space Science Institute Bern. RT acknowledges the interest of the ISSI directorate as well as the generous hospitality of the ISSI staff, in particular the assistance of the librarians Andrea Fischer and Irmela Schweitzer, and the Systems Administrator Saliba F. Saliba. We acknowledge valuable discussions with M. Leubner, R. Nakamura, and Z. V\"or\"os.


\begin{thebibliography}{99}

\bibitem[Aitchison \& Hey(1993)]{aitchison1993} Aitchison IJR, Hey AJG (1993) Gauge theories in particle physics, IOP Publishing, Bristol.


\bibitem[Amit(1984)]{amit1984} Amit DJ (1984) Field theory, the renormalization group, and critical phenomena. World Scientific, Singapore. 

\bibitem[Anderson(1963)]{anderson1963} Anderson PW (1963) Plasmons, Gauge Invariance, and Mass, Physical Review 130, 439--442, https://doi.org/10.1103/PhysRev.130.439.

\bibitem[Bailin \& Love(1993)]{bailin1993} Bailin D, Love A (1993) Introduction to gauge field theory, IOP Publishing, Bristol.

\bibitem[Balogh \& Treumann(2013)]{balogh2013} Balogh A, Treumann RA (2013) Physics of collisionless shocks, Springer Media, New York, https://doi.org/10.1007/978-1-4614-6099-2.

\bibitem[Binney et al.(1999)]{binney1999} Binney JJ, Dowrick NJ, Fisher AJ, Newman MEJ (1999) The theory of critical phenomena, Clarendon Press, Oxford.

\bibitem[Binsack(1966)]{binsack1966} Binsack JH (1966)  Plasma studies with the Imp-2 satellite, pp. 200. (unpublished) PhD-Thesis, MIT, Boston.

\bibitem[Christon et al.(1991)]{christon1991} Christon SP, Williams DJ, Mitchell DG, Huang CY, Frank LA (1991) Spectral characteristicds of plasma sheet ion and electron populations during disturbed geomagnetic conditions, Journal of Geophysical Research 96, 1-22, https://doi.org/10.1029/90JA01633.

\bibitem[Christon et al.(1989)]{christon1989} Christon SP, Williams DJ, Mitchell DG, Frank LA, Huang CY (1989) Spectral characteristicds of plasma sheet ion and electron populations during undisturbed geomagnetic conditions, Journal of Geophysical Research 94, 13409-13424, https://doi.org/10.1029/JA094iA19p13409.

\bibitem[Christon et al.(1988)]{christon1988} Christon SP, Mitchell, DG, Williams DJ, Frank LA, Huang CY, Eastman TE (1988) Energy spectra of plasma sheet ions and electrons from $\sim$ 50 eV/e to $\sim$ 1 MeV during plasma temperature transitions, Journal of Geophysical Research 93, 2562--2572, https://doi.org/10.1029/JA093iA04p02562.

\bibitem[Domenech-Garret et al.(2015)]{domenech-garret2015} Domenech-Garret JL, Tierno SP,  Conde L (2015) Non-equilibrium thermionic electron emission for metals at high temperatures, Journal of Applied Physics 118, 074904, https://doi.org/10.1063/1.49929150.

\bibitem[Eastwood et al.(2005)]{eastwood2005} Eastwood JP, Lucek EA, Mazelle C, Meziane K, Narita Y, Pickett J, Treumann RA (2005) The Foreshock, Space Science Reviews 118, 41-94, https://doi.org/10.1007/s11214-005-3824-3.

\bibitem[Einstein(1905)]{einstein1905} Einstein A (1905) \"Uber die von der molekularkinetischen Theorie der W\"arme geforderte Bewegung von in ruhenden Fl\"ussigkeiten suspendierten Teilchen, Annalen der Physik 322, 549-560, https://doi.org/10.1002/andp.19053220806.

\bibitem[Fichtner et al.(2020)]{fichtner2020} Fichtner H, Kleimann J, Yoon PH, Scherer K, Oughton S, Engelbrecht NE (2020)  On the Generation of Compressible Mirror-mode Fluctuations in the Inner Heliosheath, Astrophysical Journal 901, 76, https://doi.org/10.3847/1538-4357/abaf52.

\bibitem[Hasegawa et al.(1985)]{hasegawa1985} Hasegawa  A, Mima K,  Duong-van M (1985) Plasma distribution function in a superthermal radiation field, Physical Review Letters 54, 2608--2610, https://doi.org/10.1103/PhyRevLett.54.2608.

\bibitem[Kittel \& Kroemer(1980)]{kittel1980} Kittel C, Kroemer H (1980) Thermal physics, W. H . Freeman, New York.

\bibitem[Kolb \& Turner(1989)]{kolb1989} Kolb EW, Turner MS (1989) The early universe, Addison-Wesley, Redwood City.


\bibitem[Lazar et al.(2020)]{lazar2020} Lazar M, Scherer K, Fichtner H, Pierrard V (2020) Toward a realistic macroscopic parametrization of space plasmas with regularized $\kappa$-distribution, Astronomy and Astrophysics 643, A20, https://doi.org/10.1051/0004-6361/201936861.

\bibitem[Livadiotis \& McComas(2013)]{livadiotis2013} Livadiotis G, McComas DJ (2013) Understanding kappa distributions: A toolbox for space science and astrophysics, Space Science Reviews 175, 183--214, https://doi.org/10.1007/s11214-013-9982-9.

\bibitem[Lucek et al.(2005)]{lucek2005} Lucek EA, Constantinescu D, Goldstein ML, Pickett J, Pinc{c}on JL, Sahraoui F, Treumann RA, Walker SN (2005) The magnetosheath, Space Science Reviews 118, 95-112, https://doi.org/10.1007/s11214-005-3825-2.


\bibitem[Nambu(1960)]{nambu1960}Nambu Y (1960) Quasi-Particles and Gauge Invariance in the Theory of Superconductivity, Physical Review 117, 648-663, https://doi.org/10.1103/PhysRev.117.648.

\bibitem[Nambu \& Jona-Lasinio(1961)]{nambu1961} Nambu Y, Jona-Lasinio G (1961) Dynamical Model of Elementary Particles Based on an Analogy with Superconductivity. I \& II, Physical Review 122, 345-358, https://doi.org/10.1103/PhysRev.117.648, 124, 246-254, https://doi.org/10.1103/PhysRev.124.246.


\bibitem[Olbert(1967)]{olbert1967} Olbert S (1968) Summary of experimental results from M.I.T. detector on IMP-1, in: Physics of the Magnetosphere, Proceeding of a Conference at Boston College, June 19-28, 1967 (RDL Carovillano, JF McClay, editors), Astrophysics and Space Science Library 40, p. 641, Reidel Publ., Dordrecht.

\bibitem[Parisi(1988)]{parisi1988} Parisi G (1988) Statistical field theory, Addison-Wesley, Redwood City.


\bibitem[Reichl(1980)]{reichl1980} Reichl LE (1980) A modern course in statistical physics, Arnold Publishers, Austin TX.

\bibitem[Renyi(1970)]{renyi1970} Renyi A (1970) Probability theory, North-Holland Publishing Company.

\bibitem[Ryder(1991)]{ryder1991} Ryder LH (1991) Quantum field theory, Cambridge University Press, Cambridge.

\bibitem[Scherer et al.(2017)]{scherer2017} Scherer K, Fichtner H, Lazar M (2017) Regularized $\kappa$-distributions with non-diverging moments, Europhysics Letters EPL 120, 50002, https://doi.org/10.1209/0295-5075/120/50002.

\bibitem[Scherer et al.(2020)]{scherer2020} {Scherer K, Husidic E, Lazar M, Fichtner H (2020) The $\kappa$-cookbook: a novel generalizing approach to unify $\kappa$-like distributions for plasma particle modelling, Monthly Notices of the Royal Astronomical Society 497, 1738-1756, https://doi.org/10.1093/mnras/staa1969}

\bibitem[Treumann(1999)]{treumann1999} Treumann RA (1999) Kinetic theoretical foundation of Lorentzian statistical mechanics, Physica Scripta 59, 19-26, https://doi.org/10.1238/Physica.Regular.059a00019.

\bibitem[Treumann \& Jaroschek(2008)]{treumann2008} Treumann RA, Jaroschek CH (2008) Gibbsian theory of power law distributions, Physical Review Letters 100, 155005, https://doi.org/10.1103/PhysRevLett.100.155005.

\bibitem[Treumann \& Baumjohann(2014)]{treumann2014} Treumann RA, Baumjohann W (2014) Beyond Gibbs.Boltzmann-Shannon: General entropies -- The Gibbs-Lorentzian example, Frontiers in Physics 2, 49, https://doi.org/110.3389/fphys.2014.00049.


\bibitem[Treumann \& Baumjohann(2018)]{treumann2018} Treumann RA, Baumjohann W (2018) The differential cosmic ray energy flux in the light of an ultrarelativistic generalized Lorentzian thermodynamics, Astrophysics and Space Science 363, 37, https://doi.org/10.1007/s10509-018-3255-8.

\bibitem[Treumann \& Baumjohann(2020)]{treumann2020} Treumann RA, Baumjohann W (2020) Lorentzian entropies and Olbert's $\kappa$-distribution, Frontiers in Physics 8, 221, https://doi.org/10.3389/fphy.2020.00221.

\bibitem[Treumann et al.(2004)]{treumann2004} Treumann RA, Jaroschek CH, Scholer M (2004) Stationary plasma states far from equilibrium, Physics of Plasmas 11, 1317-1325, https://doi.org/10.1063/1.1667498.

\bibitem[Tsallis(1988)]{tsallis1988} Tsallis C (1988) Possible generalization of Boltzmann-Gibbs statistics, Journal of Statistical Physics 52, 479-487, https://doi.org/10.1007/BF01016429.

\bibitem[Vasyliunas(1968)]{vasyliunas1968} Vasyliunas VM (1968) A survey of low-energy electrons in the evening sector of the magnetosphere with OGO 1 and OGO 3, Journal of Geophysical Research 73, 2839-2884, https://doi.org/10.1029/JA073i009p02839.

\bibitem[Wehrl(1978)]{wehrl1978} Wehrl A (1978) General properties of entropy, Reviews of Modern Physics 50, 221-260, https://doi.org/10.1103/RevModPhys.50.221.

\bibitem[Wilson \& Kogut(1974)]{kogut1974} {Wilson KG, Kogut J (1974) The renormalization group and the $\epsilon$ expansion, Physics Reports 12C, 75-199, https://doi.org/10.1016/0370-1573(74(90023-4.}


\bibitem[Yoon et al.(2005)]{yoon2005} Yoon PH, Rhee T, Ryu CM (2005)  Self-consistent generation of superthermal electrons by beam-plasma interaction, Physical Review Letters 95, 215003, https://doi.org/10.1103/PhysRevLett.95.215003.

\bibitem[Yoon et al.(2006)]{yoon2006} Yoon PH, Rhee T, Ryu CM (2006)  Self-consistent generation of electron $\kappa$ distribution: 1. Theory, Journal of Geophysical Research 111, A09106, https://doi.org/10.1029/2006JA011681.

\bibitem[Yoon et al.(2018)]{yoon2018} Yoon PH, Lazar M, Scherer K, Fichtner H, Schlickeiser R (2018)  Modified ?-distribution of Solar Wind Electrons and Steady-state Langmuir Turbulence, Astrophysical Journal 868 111, 131, https://doi.org/10.3847/1538-4357/aaeb94 .

\bibitem[Zinn-Justin(1989)]{zinn1989} {Zinn-Justin J (1989) Quantum field theory and critical phenomena, Oxford University Press, Oxford.}


\end{thebibliography}
\end{document}